\documentclass[aps,prl,twocolumn,superscriptaddress,showpacs,amsmath,amssymb,tight]{revtex4-1}

\usepackage{graphicx}
\usepackage{dcolumn}
\usepackage{bm}
\usepackage{hyperref}

\begin{document}

\newcommand{\hc}{$H^{\parallel c}_{c2}$}
\newcommand{\hab}{$H^{\bot c}_{c2}$}

\title{Anisotropic upper critical field and a possible Fulde-Ferrel-Larkin-Ovchinnikov \\ state in a stoichiometric pnictide superconductor LiFeAs}

\author{K.~Cho}
\affiliation{The Ames Laboratory, Ames, IA 50011, USA}

\author{H.~Kim}
\affiliation{The Ames Laboratory, Ames, IA 50011, USA}
\affiliation{Department of Physics \& Astronomy, Iowa State University, Ames, IA 50011, USA}

\author{M.~A.~Tanatar}
\affiliation{The Ames Laboratory, Ames, IA 50011, USA}

\author{Y.~J.~Song}
\affiliation{Department of Physics, Sungkyunkwan University, Suwon, Gyeonggi-Do 440-746, Republic of Korea}

\author{Y.~S.~Kwon}
\affiliation{Department of Physics, Sungkyunkwan University, Suwon, Gyeonggi-Do 440-746, Republic of Korea}

\author{W.~A.~Coniglio}
\affiliation{Department of Physics, Clark University, Worcester, MA 01610, USA}

\author{C.~C.~Agosta}
\affiliation{Department of Physics, Clark University, Worcester, MA 01610, USA}

\author{A.~Gurevich}
\affiliation{ National High
Magnetic Field Laboratory, Florida State University, Tallahhassee, FL 32310, USA }

\author{R.~Prozorov}
\email[Corresponding author: ]{prozorov@ameslab.gov}
\affiliation{The Ames Laboratory, Ames, IA 50011, USA}
\affiliation{Department of Physics \& Astronomy, Iowa State University, Ames, IA 50011, USA}

\date{24 November 2010}

\begin{abstract}
Measurements of the temperature and angular dependencies of the upper critical field $H_{c2}$ of a stoichiometric single crystal LiFeAs in pulsed magnetic fields up to 50~T were performed using a tunnel diode resonator. Complete \hc$(T)$ and \hab$(T)$ functions with \hc$(0)=17 \pm 1$~T, \hab$(0)=26 \pm 1$~T, and the anisotropy parameter ${\gamma _H }(T) \equiv H_{c2}^{ \bot c}/H_{c2}^{\parallel c}$ decreasing from 2.5 at $T_c$ to 1.5 at $T\ll T_c$  were obtained. The results for both orientations are in excellent agreement with a theory of $H_{c2}$ for two-band $s^\pm$ pairing in the clean limit. We show that \hc$(T)$ is mostly limited by the orbital pairbreaking, whereas the shape of \hab$(T)$ indicates strong paramagnetic Pauli limiting and the inhomogeneous Fulde-Ferrel-Larkin-Ovchinnikov (FFLO)state below $T_F \sim 5$~K.
\end{abstract}

\maketitle

There are only few stoichiometric iron-based compounds (Fe-SCs) exhibiting ambient-pressure superconductivity without doping. Among those  LiFeAs is unique because of its relatively high $T_c=18$~K, \cite{Wang2008Solid} as compared to  LaFePO ($T_c=5.6$~K) \cite{Kamihara} and KFe$_{2}$As$_{2}$ ($T_c=3$~K) \cite{Rotter}. The absence of doping-induced disorder leads to weak electron scattering, low resistivity, $\rho(T_c) \approx 10 ~\mu \Omega$cm \cite{Kim} and high resistivity ratio, $RRR=\rho(300\mathrm{ K})/\rho(T_c) > 30$ \cite{Song,Kim}. These parameters differ significantly from those of most Fe-SCs for which superconductivity is induced by doping, for example, Ba(Fe$_{1-x}$T$_{x}$)$_{2}$As$_{2}$ \cite{Ni2010,Tanatar},(Ba$_{1-x}$K$_{x}$)Fe$_{2}$As$_{2}$
\cite{Rotter} and BaFe$_{2}$(As$_{1-x}$P$_{x}$)$_{2}$ \cite{Kasahara}. With the highest $T_c$ among stoichiometric Fe-SCs, negative $dT_c/dP$ \cite{Chu}, tetragonal crystal structure \cite{Wang2008Solid,Song} and the absence of antiferromagnetism \cite{Borisenko}, LiFeAs serves as a model of clean, nearly  optimally-doped Fe-SC \cite{Kim}. Because of very high $H_{c2}$ of Fe-SCs, they may also exhibit exotic behavior caused by strong magnetic fields, for example, the Fulde-Ferrel-Larkin-Ovchinnikov (FFLO) state in which the Zeeman splitting results in oscillations of the order parameter along the field direction \cite{FFLO}. Thus, measurements of $H_{c2}(T)$ in  stoichiometric LiFeAs single crystals can reveal manifestations of $s^\pm$ pairing in the clean limit \cite{Theory} for which the FFLO state would be least suppressed by doping-induced disorder \cite{FFLO} as compared to other optimally doped Fe-SCs.

Measurements of the upper critical fields parallel (\hc) and perpendicular (\hab) to the crystallographic $c-$axis in many Fe-Sc have shown several common trends \cite{Hunte,Jia,Kacmarcik,LeeBartkowiak,Jaroszynski,Kano, Bukowski,Yuan,Jiang,Butch,Ni2010,Tanatar,Braithwaite,Lei,Kida,Khim,Fang}.
Close to $T_c$ where $H_{c2}$ is limited by orbital pairbreaking, the anisotropy parameter $\gamma _H \equiv H_{c2}^{\bot c}/H_{c2}^{\parallel c}$ ranges between 1.5 and 5 \cite{Kano,Braithwaite,Kida,Lei,Khim,Hunte}, in agreement with the anisotropy of the normal state resistivity $\gamma_H = (\rho_c/\rho_{ab})^{1/2}$ above $T_c$ \cite{Tanatar}. As $T$ decreases, $H_{c2}(T)$ becomes more isotropic \cite{Yuan,Fang,Kano}, consistent with multiband pairing scenarios and the behavior of $H_{c2}$ in dirty MgB$_2$  \cite{Gurevich2003}, yet opposite to clean $s^{++}$ MgB$_2$ single crystals \cite{Kogan}. However, the more isotropic $H_{c2}$ at low $T$ can also result from strong Pauli pairbreaking for ${\bf H}\|ab$ since the observed $H_{c2}$ on many Fe-SCs significantly exceeds the BCS paramagnetic limit $H_p(T)=1.84T_c(K)$ \cite{Jaroszynski,Kano,Khim,Kida,Fang,p}. Thus, measuring $H_{c2}$ in LiFeAs can probe the interplay of orbital and Pauli pairbreaking in the clean $s^\pm$ pairing limit at high magnetic fields. These measurements are also interesting because magnetic fluctuations may contain significant ferromagnetic contribution which may lead to  triplet pairing \cite{Rice}. Experimentally, vortex properties of LiFeAs were found to be very similar to the supposedly triplet Sr$_2$RuO$_4$ \cite{Pramanik}, although NMR studies suggest singlet pairing \cite{Li}. Triplet superconductors can exhibit unusual response to magnetic field \cite{Lebed}, and, indeed, candidate materials show pronounced anomalies, as observed in UPt$_3$ \cite{Suderow,Hasselbach} and Sr$_2$RuO$_4$ \cite{Deguchi}. Surprisingly, our measurements show that normalized \hab~ of LiFeAs matches quite closely that of Sr$_2$RuO$_4$.

We present the measurements of the complete $H-T$ phase diagram of LiFeAs in pulsed magnetic fields up to 50~T, and down to 0.6~K using a tunnel diode resonator (TDR) technique. We found that \hab$(T)$ shows rapid saturation at low temperatures, consistent with strong Pauli pairbreaking. Similar conclusion was reached from torque measurements \cite{Kurita}. Our data  can be described well by a theory of $H_{c2}$ for the multiband $s^\pm$ pairing in the clean limit \cite{Gurevich2010}, which also suggests the FFLO state in LiFeAs for ${H}\bot c$ below 5~K.  Previous measurements of $H_{c2}$ in LiFeAs were performed at relatively low fields \cite{Song, LeeKhim}, thus not allowing to reveal the  spin-limited behavior at low $T$. The only reported high-field measurements associate $H_{c2}$ with the disappearance of irreversibility in torque measurements Ref.~\cite{Kurita}. The authors supported this association by comparing with the specific heat data. However, in our opinion, the irreversibility field may underestimate the true $H_{c2}(T)$ and have different temperature dependence due to depinning of vortices. It may also have significant (cusp like) angular variation, which would be particularly important for torque measurements that rely on the finite angle between magnetic moment and field. Related complications were discussed in high$-T_c$ cuprates \cite{Carrington}.

\begin{figure}[tb]
\includegraphics[width=8.5cm]{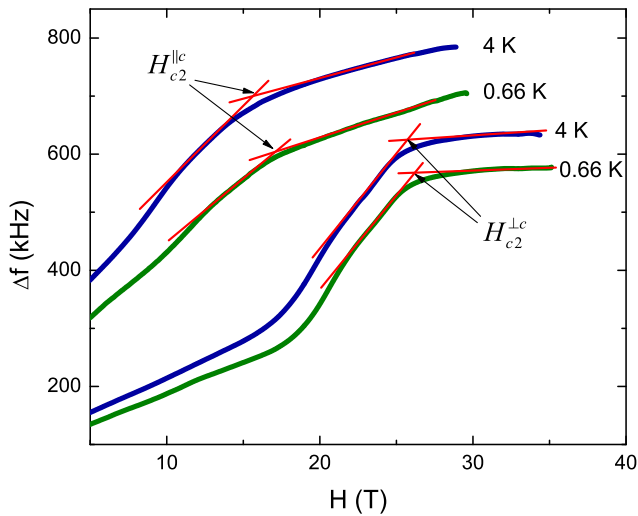}
\caption{\label{fig:fig1} (Color online) TDR frequency change for increasing pulsed magnetic field, appied in two orientations, $H \parallel c$ and $H \bot c$, shown for two temperatures for sample A. The definition of $H_{c2}$ is shown as the intersection of two straight lines below and above the transition.}
\end{figure}

\begin{figure}[tb]
\includegraphics[width=8.5cm]{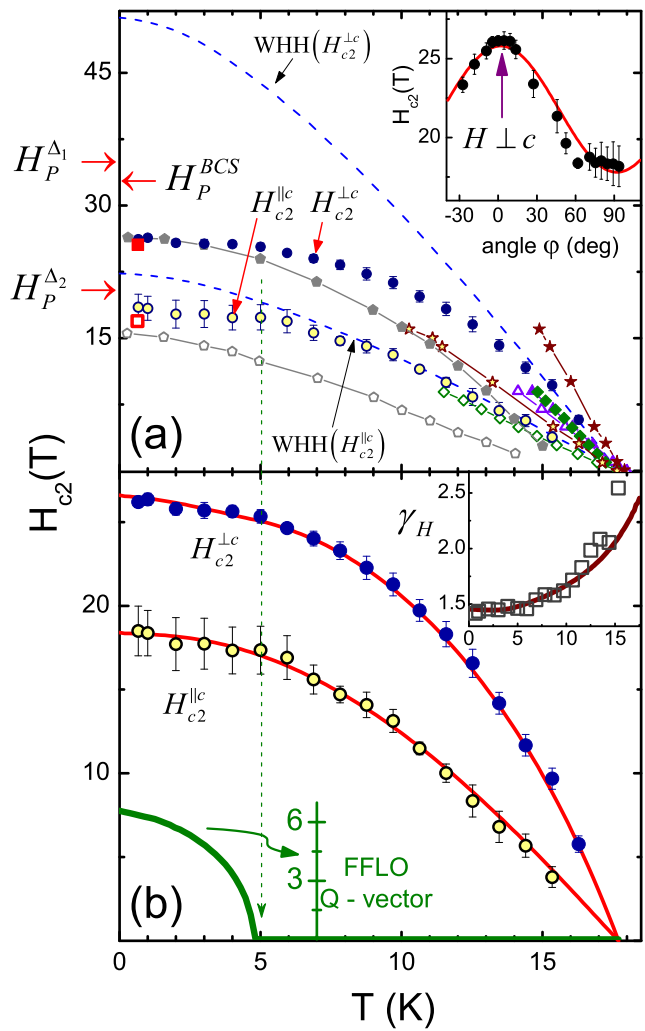}
\caption{\label{fig:fig2} (Color online) (a) $H_{c2}(T)$ for $H \bot c$ (solid symbols) and $H \parallel c$ (open symbols). Blue circles and red squares correspond to samples A and B, respectively. For comparison we show the literature data determined from the resistivity measurements with mid-point criterion: (magenta) triangles \cite{Song}, (green) rhombi  \cite{LeeKhim}, (brown) stars \cite{Heyer}. Torque data are shown by (grey) pentagons \cite{Kurita}. Dashed lines is the WHH $H_{c2}(T)$. Inset in (a) shows $H_{c2}(\varphi)$ at $0.66$~K where the solid line is $H_{c2}(\varphi) =H_{c2}^{\| c}+(H_{c2}^{\bot c}-H_{c2}^{\| c}) \cos\varphi$. (b) Fit of the experimental data to  $H_{c2}(T)$, $Q(T)$ and $\gamma_H(T)$ (solid lines) calculated from Eq.~(\ref{mgh}) for the parameters given in the text. The FFLO wave vector $Q(T)$ is plotted in the units of $40\pi k_BT_cg_1/\hbar v_1 $, and the inset shows the anisotropy parameter $\gamma _{H}(T)$. }
\end{figure}

Single crystals of LiFeAs were grown in a sealed tungsten crucible using Bridgeman method and placed in ampoules. Immediately after opening, samples were covered with Apiezon N grease, which provides some degree of short-term protection \cite{Kim}. The samples were cleaved and cut inside the grease layer to minimize exposure to the air. The two studied samples had dimensions of $0.6 \times 0.5 \times 0.1$ mm$^{3}$ (sample A) and $0.9 \times 0.8 \times 0.2$ mm$^{3}$ (sample B). Superconducting transition temperature for both samples was $T_c = 17.6\pm0.1$~K (more than 10\% higher than $T_c = 15.5$~K of Ref.~\onlinecite{Kurita}). Dynamic magnetic susceptibility $\chi$ was measured with 190 MHz (sample A) and 16 MHz (sample B) TDR \cite{Prozorov2006}. The magnetic field was generated by a 50 T pulsed magnet with a 11 ms rise time at Clark University. A single-axis rotator with a 0.5$^{\circ}$ angular resolution was used to accurately align the sample with respect to the $c-$axis (see inset in Fig.~\ref{fig:fig2}(a)). The data have been taken for each orientation at temperatures down to 0.66 K. The normal state data at 25 K have also been taken for both orientations and subtracted. Measured shift of the resonant frequency $\Delta f \propto \chi$ \cite{Prozorov2006}, thus exhibits a kink at $H_{c2}$ where London penetration depth diverges and is replaced by the normal - state skin depth. Thus, barring uncertainty due to fluctuations, it is probing a ``true'' upper critical field.

Fig.~\ref{fig:fig1} shows the change of the resonant frequency as a function
of $H$ for sample A for two field orientations and two temperatures. From many such traces, both  \hab~ and \hc~
were determined as shown in Fig.~\ref{fig:fig1} and are plotted in Fig.~\ref{fig:fig2}. Figure~\ref{fig:fig2}(a) compares our $H_{c2}$ data
on samples A and B with the previous transport \cite{Song, LeeKhim, Heyer} and torque measurements \cite{Kurita}. Figure~\ref{fig:fig2}(a) also shows the behavior expected from the orbital Werthamer-Helfand-Hohenberg (WHH) theory \cite{Werthamer} with $H_{\rm orb}(0)=0.69T_{\rm c}|dH_{\rm c2}/dT|_{\rm T_{\rm c}}$, the single-gap BCS paramagnetic limit, $H^{BCS}_P = 1.84 T_c=32.2$~T, as well as $H^{\Delta_1}_P=34.7$~T and $H^{\Delta_2}_P=20.4$~T calculated with $\Delta _1$(0)/T$_c \approx 1.885$ and $\Delta _1$(0)/T$_c \approx 1.111$ reported for the same samples in Ref.~\cite{Kim}. Clearly, the observed $H_{c2}(T)$ exhibits much stronger flattening at low temperature compared to the orbital WHH theory. Inset in Fig.~\ref{fig:fig2}(a) shows the dependence of $H_{c2}$ on the angle between ${\bf H}$ and the $ab$ plane at 0.66~K where $H_{c2}^{ \bot c}$ is defined at a maximum of $H_{c2}(\varphi) =H_{c2}^{\| c}+(H_{c2}^{\bot c}-H_{c2}^{\| c}) \cos\varphi$ depicted by the solid line.

We analyze our $H_{c2}(T)$ data using a two-band theory, which takes into account both orbital and paramagnetic pairbreaking in the clean limit, and the possibility of the FFLO with the wave vector $Q(T,H)$. In this case the equation for $H_{c2}$ is given by \cite{Gurevich2010},

\begin{gather}
    a_1G_1+a_2G_2+G_1G_2=0,
\label{mgh} \\
    G_1=\ln t+2e^{q^{2}}\operatorname{Re}\sum_{n=0}^{\infty}\int_{q}^{\infty}due^{-u^{2}}\times
\nonumber \\
    \left[\frac{u}{n+1/2}-\frac{t}{\sqrt{b}}\tan^{-1}\left(
    \frac{u\sqrt{b}}{t(n+1/2)+i\alpha b}\right)\right].
\label{U1}
\end{gather}

 \noindent Here $Q(T,H)$ is determined by the condition that $H_{c2}(T,Q)$ is maximum, $a_1=(\lambda_0 +\lambda_{-})/2w$, $a_2=(\lambda_0-\lambda_{-})/2w$, $\lambda_{-}=\lambda_{11}-\lambda_{22}$, $\lambda_0=(\lambda_{-}^2+4\lambda_{12}\lambda_{21})^{1/2}$, $w=\lambda_{11}\lambda_{22}-\lambda_{12}\lambda_{21}$, $t=T/T_c$, and $G_2$ is obtained by replacing $\sqrt{b}\to\sqrt{\eta b}$ and $q\to q\sqrt{s}$ in $G_1$, where

\begin{gather}
    b=\frac{\hbar^{2}v^{2}_1 H }{8\pi\phi_{0}k_B^2T_c^2g_1^2},\qquad\alpha=\frac{4\mu\phi_{0}g_1k_BT_{c}}{\hbar^{2}v^2_1},
    \label{parm1} \\
    q^{2}=Q_{z}^{2}\phi_{0}\epsilon_1/2\pi H, \qquad \eta = v_2^2/v_1^2, \qquad
    s=\epsilon_2/\epsilon_1.
    \label{parm2}
\end{gather}

\noindent Here $v_l$ is the in-plane Fermi velocity in band $l={1,2}$, $\epsilon_{l}=m_{l}^{ab}/m_{l}^c$ is the mass anisotropy ratio, $\phi_0$ is the flux quantum, $\mu$ is the magnetic moment of a quasiparticle, $\lambda_{11}$ and $\lambda_{22}$ are the intraband pairing constants, and $\lambda_{12}$ and $\lambda_{21}$ are the interband pairing constants, and $\alpha\approx 0.56\alpha_M$ where
the Maki parameter $\alpha_M=H_{c2}^{orb}/\sqrt{2}H_p$ quantifies the strength of the Zeeman pairbreaking.  The factors $g_1=1+\lambda_{11}+|\lambda_{12}$ and $g_2=1+\lambda_{22}+|\lambda_{21}|$ describe the strong coupling Eliashberg corrections. For the sake of simplicity, we consider
here the case of $\epsilon_1=\epsilon_2=\epsilon$ for which \hab~ is defined by Eqs.~ (\ref{mgh}) and (\ref{U1}) with $g_1=g_2$ and rescaled $q\to q\epsilon^{-3/4}$, $\alpha\to
\alpha\epsilon^{-1/2} $ and $\sqrt{b}\to \epsilon^{1/4}\sqrt{b}$ in $G_1$ and $\sqrt{\eta b}\to \epsilon^{1/4}\sqrt{\eta b}$ in $G_2$ \cite{Gurevich2010}.

Figure~\ref{fig:fig2}(b) shows the fit of the measured $H_{c2}(T)$ to Eq.~(\ref{mgh}) for $s^{\pm}$ pairing with $\lambda_{11}=\lambda_{22}=0$, $\lambda_{12}\lambda_{21}=0.25$,  $\eta = 0.3$, $\alpha=0.35$, and $\epsilon = 0.128$. Equation (\ref{mgh}) describes \hc (T), \hab (T)~ and $\gamma _H (T)=b_\|(T)/\sqrt{\epsilon}b_\bot(T)$ where $b_\|(T)$ and $b_\bot(T)$ are the solutions of Eq.~(\ref{mgh}) for $H \|c$ and $H \bot c$, very well. The fit parameters are also in good quantitative agreement with experiment. For instance, the Fermi velocity $v_1=(g_1k_BT_c/\hbar)[8\pi\phi_0b_\bot(0)/H^{\parallel c}_{c2}(0)]^{1/2}$ can be expressed  from Eq.~(\ref{parm2}) in terms of materials parameters and $b_\bot(0)=0.314$ calculated from Eq.~(\ref{mgh}). For $T_c=17.8$~K, $H^{\parallel c}_{c2}(0)=18.4$T and $g=1.5$ for $\lambda_{12}=0.5$, we obtain $v_1=1.12\times 10^7$ cm/s, consistent with the ARPES results \cite{Borisenko}.

Several important conclusions follow from the results shown in Fig.~\ref{fig:fig2}(b). First, contrary to the single-band Ginzburg-Landau scaling, $\gamma_H^{GL}=\epsilon^{-1/2}$, the anisotropy parameter $\gamma_H(T)$ decreases as $T$ decreases. Not only is this behavior indicative of multiband pairing \cite{Gurevich2003}, but it also reflects the significant role of the Zeeman pairbreaking
in LiFeAs given that $\alpha_\| = \alpha/\sqrt{\epsilon}=0.98$ for $H \bot c$ is close to
the single-band FFLO instability threshold, $\alpha\approx 1$ \cite{Gurevich2010}. In this case $\gamma_H (T)$ near $T_c$ is determined by the orbital pairbreaking and the mass anisotropy $\epsilon$, but as $T$ decreases, the contribution of the isotropic paramagnetic pairbreaking increases, resulting in the decrease of $\gamma_H(T)$. Another intriguing result is that the solution of Eq.~(\ref{mgh}) shows no FFLO instability for $H \|c$, but predicts the FFLO transition at $T< T_F\approx 5$ K for $H ||ab$.  The FFLO wave vector $Q(T)=4\pi k_BT_cq(T)b^{1/2}(T)g_1/hv_1$  appears spontaneously at $T=T_F \approx 5$ K where the FFLO period $\ell = 2\pi/Q=\hbar v_1/2k_BT_cg_1q(T)b^{1/2}(T)$ diverges and then decreases as $T$ decreases, reaching  $\ell(0)=\pi\xi_0/g_1q(0)b^{1/2}(0)\approx 9 \xi_0$ at $T=0$. Here $q(0)=0.656$, $b(0)=0.126$, and $\xi_0=\hbar v_1/2\pi k_BT_c\simeq 7.3$ nm, giving $\ell(0)\simeq 65.6$ nm for the parameters used above. The period  $\ell(0)$ is much smaller than the mean free path, $\ell_{mfp} \sim 550$ nm, estimated from the Drude formula for an ellipsoidal Fermi surface with $\epsilon = 0.128$, $v_F=112$~km/s, $m_{ab}$ equal to the free electron mass, and $\rho(T_c)=10\mu\Omega$cm. Notice that $\rho(T_c)$ may contain a significant contribution from inelastic scattering, so the mean free path for elastic impurity scattering which destroys the FFLO state \cite{FFLO} is even larger than $\ell_{mfp}$. Therefore, the FFLO state found in our calculations may be a realistic possibility verifiable by specific heat, magnetic torque and thermal conductivity measurements.

\begin{figure}[tb]
\includegraphics[width=8.5cm]{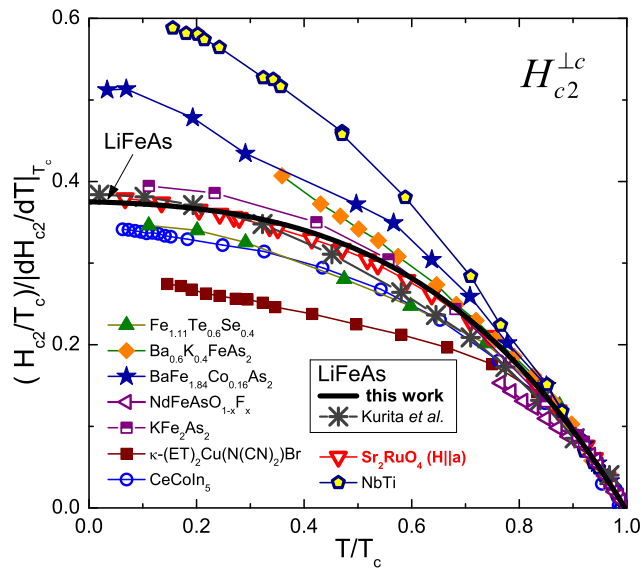}
\caption{\label{fig:fig3} (Color online) $(H_{c2}(T)/T_c)/|dH_{c2}/dT|_{\rm T_c}$ vs. $T/T_c$ in the $H \bot c$ orientation. Black solid line is our data in comparison with several Fe-SCs as well as other exotic superconductors and conventional NbTi, all shown in the legend.}
\end{figure}

Finally, we compare LiFeAs with other superconductors, especially those for which $H_{c2}$ is clearly limited by either orbital or Zeeman pairbreaking mechanisms. Shown in Fig.~\ref{fig:fig3} are the plots of the normalized $H_{c2}(T)/T_cH_{c2}'$ as functions of $T/T_c$  for the ${\bf H}\| ab$ orientation where the Zeeman pairbreaking is most pronounced. Here $H_{c2}'=|dH_{c2}/dT|_{T\to T_c}$ and our data are shown by the thick solid black line, whereas the literature data are shown by symbols. The reference materials include $H_{c2}$ for: LiFeAs \cite{Kurita}; Pauli-limited \cite{Kovalev} organic superconductor $\kappa$-(BEDT-TTF)$_2$Cu[N(CN)$_2$]Br \cite{Ohmichi}; heavy fermion CeCoIn$_5$ \cite{Bianchi}; optimally-doped iron pnictides, Ba(Fe$_{1-x}$Co$_x$)$_2$As$_2$ \cite{Kano} and  Ba$_x$K$_{1-x}$FeAs$_2$ \cite{Yuan} as well as iron chalcogenide Fe(Se,Te) \cite{Fang}. Conventional NbTi is also shown by open pentagons \cite{Shapira}. Remarkably, scaled data obtained on crystals with different $T_c$s and by different measurements (this work and Ref.~\cite{Kurita}) are very similar indicating intrinsic behavior of LiFeAs, namely, that it is indeed closer to the paramagnetic limit. Notably, the data for LiFeAs lay below other Fe-SCs, except for the highest purity ($RRR \approx 87$) KFe$_2$As$_2$ \cite{Terashima}. On the other hand, our data appear above CeCoIn$_5$, believed to be mostly Pauli limited \cite{Bianchi}. Interestingly, the data for LiFeAs stay almost on top of the $H_{c2}(T)$ for Sr$_2$RuO$_4$, in which limiting of $H_{c2}$ proceeds in a very unusual manner, leading to the formation of the second superconducting phase \cite{Deguchi}. Given that vortex dynamics in these two materials is also similar \cite{Pramanik}, the coincidence of the $H_{c2}(T)/T_cH_{c2}'$ curves is worth of further exploration.

Summarizing, full - temperature range experimental \hc$(T)$ and \hab$(T)$ deviate significantly from the single-band WHH behavior but are in excellent agreement with the theory of $H_{c2}$ for the $s^\pm$ pairing in the clean limit. Our results indicate Pauli-limited behavior and the FFLO state below 5 K for $H\bot c$.

\begin{acknowledgments}
We thank A. Carrington, V. G. Kogan, L. Taillefer and T. Terashima for discussions. The work at Ames Laboratory was supported by the U.S. Department of Energy, Office of Basic Energy Sciences, Division of Materials Sciences and Engineering under contract No. DE-AC02-07CH11358. The work at Clark was supported by the U. S. Department of Energy under contract No. ER46214. The work at Sungkyunkwan University was supported by the Basic Science Research Program (2010-0007487), the Mid-career Researcher Program (No.R01-2008-000-20586-0). R. P. acknowledges support from Alfred P. Sloan Foundation. A.~G. was supported by NSF through NSF-DMR-0084173 and by the State of Florida.

\end{acknowledgments}

\end{document}